\documentclass[aps,showpacs,preprint,superscriptaddress]{revtex4}
\usepackage{graphicx}
\usepackage{subfigure}
\usepackage{float}
\usepackage{amsmath}
\usepackage{amsfonts}
\usepackage{bm}
\usepackage{bbm}
\usepackage{txfonts}
\usepackage{array}
\usepackage{amssymb}
\usepackage{color}

\begin{document}

\title{Schwinger pair production correction in thermal system}
\author{Y. L. Wang}
\affiliation{Key Laboratory of Beam Technology of the Ministry of Education, and College of Nuclear Science and Technology, Beijing Normal University, Beijing 100875, China}
\author{H. B. Sang}
\affiliation{Key Laboratory of Beam Technology of the Ministry of Education, and College of Nuclear Science and Technology, Beijing Normal University, Beijing 100875, China}
\author{B. S. Xie}
\email[Corresponding author. Email: ]{bsxie@bnu.edu.cn}
\affiliation{Key Laboratory of Beam Technology of the Ministry of Education, and College of Nuclear Science and Technology, Beijing Normal University, Beijing 100875, China}
\affiliation{Beijing Radiation Center, Beijing 100875, China}
\date{\today}

\begin{abstract}
In this paper, we give formal results of Schwinger pair production correction in thermal systems with external background field by using the evolution operator method of thermo field dynamics, where especially tree level correction of thermal photons is considered with linear response approaches by an effective mass shift. We consider initial systems in two types of vacuums as zero temperature and thermal vacuum, respectively, with correction of thermal photons is or not included. As an example we give results of these corrections to pair production for a constant external background electric field.
\end{abstract}

\maketitle

\section{\label{sec:level1}Introduction}

In an early work of Schwinger, it shows that a very strong constant electric field decays to real charged pairs \cite{Schwinger1951}.
The produced charged pairs are observable but not observed directly yet for the absence of strong electric field near the critical field strength $E_{c}=m^{2}_{e}c^{3}/e\hbar=1.3 \times10^{16}\mathrm{V/cm}$ in laboratories. Recently, a series of papers show that direct tests of vacuum decay are planing in the near future \cite{Gavrilov2019,Dunne2014,Piazza2012}. Besides, some calculations indicate thermal enhancement of pair production \cite{Gould2017,Monin2010,Brown2018,Draper2018,Medina2017,Korwar2018} and it is similar to dynamically assisted Schwinger pair production \cite{Torgrimsson2019}, consequently experiment on thermal assisted pair production is proposed \cite{Gould2019}.

Various methods obatin identical results at zero temperature \cite{Gelis2016} but have distinct results at finite temperature for thermal pair production studied in different mechanisms. Charged particles or external electromagnetic fields are assumed to be thermalized initially, and represented in real- or imaginary-time form. Early discussion \cite{Dittrich1979,Gies1999oneloop,Elmfors1995} combined with Schwinger's approach proved that there is no thermal contribution to imaginary part of an effective one-loop action. But recent papers using worldline instants technique, where charged particles absorb thermal photons to reach maximum decay at a specific temperature \cite{Gould2017,Monin2010,Brown2018,Draper2018,Medina2017,Korwar2018}, prove that thermal effect enhances pair production. These also indicate that there is a temperature threshold $T_{c}=qE/2\pi m$ \cite{Brown2018} and discrete resonant peaks $T_{\ast}=nqE/2\pi m$ \cite{Medina2017} in a constant electric field $E$, where $q$ and $m$ are charge and mass. Moreover, Kim's works \cite{Kim2008} using  operator method and Gavrilov'works \cite{Gavrilov2008,Gavrilov2008matrix} show that there are thermal corrections in thermal quantum electrodynamics (QED) pair production. Here we should like to emphasize the following three points. First, the Schwinger pair production is a nonequilibrium process particularly when the field strength is greater than temperature. Second, in imaginary-time form, the asymptotic structure of charged particles are changed by inserting the thermal conditions in the lagrangian, the redefinition of asymptotic states may cause counting problem \cite{Krekora2006}. Last, the relation of decay rate, effective Lagrangian and pair production number $\Gamma(A)=2\mathrm{Im}\mathcal{L}_{\mathrm{eff}}(A)=\pm \sum_{\mathbf{k}}\mathrm{ln}(1\pm \mathcal{N}_{\mathbf{k}})$ \cite{Schwinger1951} changes at finite temperature in real-time form \cite{Rose2013}. Therefore, at finite temperature, the imaginary part of effective Lagrangian may not equal to measured pair production rate. Furthermore, in general thermal QED systems, the thermal photons are present, and usually are considered in loop corrections but ignored in tree level calculations. We notice that they also influence the structure of vacuum, consequently affect the result of pair production.

Therefore to obtain a measurable pair production rate at finite temperature, and to make a distinction among thermal charged particles, thermal photons and external electromagnetic fields, we use the operator method of thermo field dynamics (TFD) adding an effective mass shift approach. The core of our paper is to combine
Bogoliubov transformation which depends on the solution of Klein-Gordon equation or Dirac equation under external fields \cite{Kim2008}, thermal transformation which depends on the choice of inequivalent thermal vacuums, and effective mass corrections which concern the effect of thermal photons. Our approaches adopt almost-independent approximation for particles, consequently all interactions among charged particles are ignored, and the single particle equations have same accuracy with one-loop effective Lagrangian \cite{Kim2008} and the high order corrections are negligible \cite{Gies2017}. In Dirac vacuum, negative energy(frequency) electron absorbs photons to become electrons, and its in- and out- states depend on background fields, which are assumed to consists of external space-time dependent field, $A^{ex}_{\mu}(x)$, and temperature dependent field, $A^{\gamma}_{\mu}(T_{r})$ so that $A_{\mu}=A^{ex}_{\mu}(x)+A^{\gamma}_{\mu}(T_{r})$. Assuming that thermal fields are turned off when $t\rightarrow\infty$ (before measurement), and employing semiclassical mass shift $m^{*2}=m^{2}-q^{2}\langle A^{\gamma,\mu}A^{\gamma}_{\mu}\rangle$ to reflect the corrections of thermal photon fields, we find that $T_{r}$ appears in the exponent which decreases the threshold of tunneling so that it enhances the rate of pair production.

The paper is organized as follows. In Sec.~II, we revisit the pair production in vacuum
then consider it in thermal charged particles, thermal photons and their mixture. We will show their similar results only with change
in thermal photons and the mixture when temperature is small $T_{r}\ll m$. In Sec.~III, we apply the method to a constant electric field and give the integral form and low-temperature polynomial form. In Sec.~IV, we summarize our results, and point out future directions. We use the natural units where $c=\hbar=k_{B}=\varepsilon_{0}=1$. Inverse of temperature is denoted as $\beta=1/T$. The charge and the mass of boson and fermion have unified marks as $q$ and $m$.

\section{\label{sec:1}QED pair production in thermal systems}

This method is based on motion equation, and it is convenient to take Dirac equation and Klein-Gordon equation into the form
\begin{equation}\label{KGDirac}
\begin{split}
\left\{
(\partial_{\mu}-iqA_{\mu})^{2}\!+\!m^{2}\!+\!
2q\left(
\begin{matrix}
(\mathbf{B}\!+\!i\mathbf{E})\cdot\mathbf{\mathbf{\sigma}_{s}}& \\ &(\mathbf{B}\!-\!i\mathbf{E})\cdot\mathbf{\mathbf{\sigma}_{s}}
\end{matrix}
\right)
\right\}& \\
\times \phi_{s}=0&,
\end{split}
\end{equation}
where $\mathbf{B}$ and $\mathbf{E}$ are magnetic and electric fields, and $\mathbf{\sigma_{s}}$ is half of pauli matrix for fermion and zero matrix for boson. In this equation,
the difference depends on their spin where spin $s=\pm1/2$ for spinor QED and spin $s=0$ for scalar QED .

\subsection{\label{sec:level2}QED pair production at zero temperature}

The annihilation operator of incoming state $\phi^{\small{(+)}}_{\mathbf{k},\mathrm{in}}$ is $a_{\mathbf{k}}(t_{\mathrm{in}}=-\infty)$ for particle
and $b_{\mathbf{k}}(t_{\mathrm{in}}=-\infty)$ for antiparticle state $\phi^{(-)}_{\mathbf{k},\mathrm{in}}$,
and similarly the annihilation operator of outgoing states $\phi^{(+)}_{\mathbf{k},\mathrm{out}}$ and $\phi^{(-)}_{\mathbf{k},\mathrm{\mathrm{out}}}$ are defined by $a_{\mathbf{k}}(t_{\mathrm{out}}=+\infty)$ and $b_{\mathbf{k}}(t_{\mathrm{out}}=+\infty)$. The subscript $\mathbf{k}$ denotes $(\mathbf{k},s)$ including momentum and spin, and the superscript $(+)$ and $(-)$ distinguish the positive energy and negative energy states. The incoming states and outgoing states depend on the gauge potential $A(x)$ that determines the asymptotic solutions of Eq.~\eqref{KGDirac}.

The in-vacuum and out-vacuum $|0,\mathrm{in}\rangle$ and $|0,\mathrm{out}\rangle$ are defined in the particle number representation as:
\begin{equation}
|0,\mathrm{in}\rangle =\prod_{\mathbf{k}}\otimes|0_{\mathbf{k}},\mathrm{in}\rangle\,
\end{equation}
and
\begin{gather}\label{EvolutionVacuum}
|0,\mathrm{out}\rangle=\prod_{\mathbf{k}}U_{\mathbf{k}}(A^{ex})|0,\mathrm{in}\rangle=U(A^{ex})|0,\mathrm{in}\rangle.
\end{gather}
The $U(A^{ex})$ is an evolution operator and satisfies
\begin{gather}
a_{\mathbf{k},\mathrm{out}}=U_{\mathbf{k}}(A^{ex}) a_{\mathbf{k},\mathrm{in}} U^{+}_{\mathbf{k}}(A^{ex}).
\end{gather}
The outgoing operators and ingoing operators are related through the Bogoliubov transformation
\begin{equation}\label{BBTT}
\begin{aligned}
& a_{\mathbf{\mathbf{k}},\mathrm{out}}=\int \frac{d^{3}\mathbf{k^{'}}}{2E_{\mathbf{k}}(2\pi)^{3}}
\left[\langle \phi^{(+)}_{\mathbf{k},\mathrm{out}}|\phi^{(+)}_{\mathbf{k}^{'},\mathrm{in}} \rangle a_{\mathbf{k}^{'}}+
\langle \phi^{(+)}_{\mathbf{k},\mathrm{out}}|\phi^{(-)}_{\mathbf{k}^{'},\mathrm{in}} \rangle b^{+}_{\mathbf{k}^{'}}\right],\\
& b^{+}_{\mathbf{\mathbf{k}},\mathrm{out}}=\int \frac{d^{3}\mathbf{k^{'}}}{2E_{\mathbf{k}}(2\pi)^{3}}
\left[\langle \phi^{(-)}_{\mathbf{k},\mathrm{out}}|\phi^{(+)}_{\mathbf{k}^{'},\mathrm{in}} \rangle a_{\mathbf{k}^{'}}+
\langle \phi^{(-)}_{\mathbf{k},\mathrm{out}}|\phi^{(-)}_{\mathbf{k}^{'},\mathrm{in}} \rangle b^{+}_{\mathbf{k}^{'}}\right].
\end{aligned}
\end{equation}
The four projections of in-basis and out-basis are Bogoliubov coefficients that can be solved numerically by sloving Eq.~\eqref{KGDirac} in iterative algorithm with fast Fourier transform (FFT) \cite{Braun1999,Blinne2019} for arbitrary fields. When Eq.~\eqref{KGDirac} has an analytical solution in a certain field, Eq.~\eqref{BBTT} has a compact form
\begin{equation}\label{ExternalBT}
\left(
\begin{aligned}
& a_{\mathbf{\mathbf{k}},\mathrm{out}} \\
& b^{+}_{\mathbf{\mathbf{k}},\mathrm{out}}
\end{aligned}
\right) =
\left(
\begin{matrix}
\mu_{\mathbf{k}} & \nu^{*}_{\mathbf{k}} \\ \eta\nu_{\mathbf{k}} & \mu^{*}_{\mathbf{k}}
\end{matrix}
\right)
\left(
\begin{aligned}
& a_{\mathbf{\mathbf{k}},\mathrm{in}} \\
& b^{+}_{\mathbf{\mathbf{k}},\mathrm{in}}
\end{aligned}
\right),
\end{equation}
where either $\eta=-1$ for boson or $\eta=1$ for fermion, and the coefficients have general form
\begin{equation}
\mu_{\mathbf{k}} =\langle \phi^{(+)}_{\mathbf{k},\mathrm{out}} |\phi^{(+)}_{\mathbf{k},\mathrm{in}} \rangle,~\\
\nu^{*}_{\mathbf{k}}=\langle \phi^{(+)}_{\mathbf{k},\mathrm{out}} |\phi^{(-)}_{\mathbf{k},\mathrm{in}} \rangle.
\end{equation}
These Bogoliubov coefficients satisfy the relation
\begin{equation}
|\mu_{\mathbf{k}}|^{2} + \eta |\nu_{\mathbf{k}} |^{2}=1.
\end{equation}

Combining transformation relation Eq.~\eqref{ExternalBT} and number operator $a^{+}_{\mathrm{in}}a_{\mathrm{in}}$ leads to the number of produced pairs with $\mathbf{k}$ momentum\footnote{It is $\langle 0,\mathrm{in}|a^{+}_{\mathbf{k},\mathrm{out}} a_{\mathbf{k},\mathrm{out}}|0,\mathrm{in} \rangle$ in some literature but has no difference in result, since the transformation is unitary.}
\begin{equation}\label{kn}
\mathcal{N}_{\mathbf{k}}
= \langle 0,\mathrm{out}|a^{+}_{\mathbf{k},\mathrm{in}} a_{\mathbf{k},\mathrm{in}}|0,\mathrm{out} \rangle = |\nu_{\mathbf{k}}|^{2}.
\end{equation}
Finally, the total number of produced pairs
\begin{equation}\label{N}
\mathcal{N}=\int \frac{d\mathbf{k}}{(2\pi)^{3}}|\nu_{\mathbf{k}}|^{2}.
\end{equation}

\subsection{\label{sec:q}QED pair production in thermal charged particles}

In thermal QED systems, real charged particles condense when spontaneous vacuum breaking occurs even under external current. Here virtual particles are ignored for
we only consider one loop. Systems are assumed to be ideal canonical ensemble, where chemical potential is ignored.
The thermal vacuum is defined as \cite{Das1997,Ojima1981}
\begin{equation}
|0,\beta_{q},\mathrm{in}\rangle=\sum_{\mathbf{k}}\sum_{n_{\mathbf{k}}}Z^{-1/2}_{\mathrm{in}} e^{-\beta_{q}n_{\mathbf{k}}\varepsilon_{\mathbf{k}}/2}|n_{\mathbf{k}},\tilde{n}_{\mathbf{k}},\mathrm{in}\rangle.
\end{equation}
where $n_{\mathbf{k}}$ and $\tilde{n}_{\mathbf{k}}$ are the number of particle corresponding to $a_{\mathbf{k},\mathrm{in}}$ and $\tilde{a}_{\mathbf{k},\mathrm{in}}$ respectively, and satisfy $n_{\mathbf{k}}=\tilde{n}_{\mathbf{k}}$. The thermal vacuum can be treated as thermal rotation of incoming vacuum that $U(\theta)|0,\mathrm{in}\rangle=|0,\beta_{q},\mathrm{in}\rangle$ where
\begin{equation}
U(\theta)=e^{-\sum_{\mathbf{k}}\theta_{\mathbf{k}}(\beta_{q})(\tilde{c}_{\mathbf{k},\mathrm{in}}c_{\mathbf{k},\mathrm{in}} - c^{+}_{\mathbf{k},\mathrm{in}}\tilde{c}^{+}_{\mathbf{k},\mathrm{in}})}.
\end{equation}
The relation between thermalized incoming particles and original incoming particles is
\begin{equation}\label{ThermalBT}
\left(
\begin{aligned}
&c_{\mathbf{k},\mathrm{in}}(\theta) \\
&\tilde{c}^{+}_{\mathbf{k},\mathrm{in}}(\theta)
\end{aligned}
\right) =
\left(
\begin{matrix}
\mathrm{cos(h)}\theta_{\mathbf{k}} & -\mathrm{sin(h)}\theta_{\mathbf{k}} \\ \eta \mathrm{sin(h)}\theta_{\mathbf{k}} &\mathrm{ cos(h)}\theta_{\mathbf{k}}
\end{matrix}
\right)
\left(
\begin{aligned}
&c_{\mathbf{k},\mathrm{in}} \\
&\tilde{c}^{+}_{\mathbf{k},\mathrm{in}}
\end{aligned}
\right),
\end{equation}
where fermion picks trigonometric function $\mathrm{sin}\theta/\mathrm{cos}\theta$ and boson picks hyperbolic function $\mathrm{sinh}\theta/\mathrm{cosh}\theta$. These are represented as
\begin{equation}\label{cossin}
\mathrm{sin(h)}\theta_{\mathbf{k}}=\frac{e^{-\beta_{q}\varepsilon_{\mathbf{k}}/2}}{\sqrt{1+\eta e^{-\beta_{q}\varepsilon_{\mathbf{k}}}}},~
\mathrm{cos(h)}\theta_{\mathbf{k}}=\frac{1}{\sqrt{1+\eta e^{-\beta_{q}\varepsilon_{\mathbf{k}}}}}.
\end{equation}
When thermal annihilation operators act on the thermal vacuum, it satisfies
\begin{equation}\label{c+c}
\begin{split}
&c_{\mathbf{k},\mathrm{in}}(\theta)|0,\beta_{q},in\rangle=\tilde{c}_{\mathbf{k},\mathrm{in}}(\theta)|0,\beta_{q},\mathrm{in}\rangle=0,\\
&c_{\mathbf{k}}|0,\mathrm{in}\rangle = \tilde{c}_{\mathbf{k}}|0,\mathrm{in}\rangle =0 .
\end{split}
\end{equation}
For all the dynamical observables are only made of non-tilde $c$ operators in TFD, the number operator is just $c^{+}c$. Now we construct a doubling thermal vacuum
\begin{equation}
\begin{split}
|0,0,\beta_{q},\mathrm{in}\rangle=\sum_{\mathbf{k}}\sum_{n_{\mathbf{k}}}\sum_{m_{\mathbf{k}}}\delta_{n_{\mathbf{k}},m_{\mathbf{k}}}
Z^{-1/2}_{\mathrm{in}} e^{-\beta_{q}(n_{\mathbf{k}}+m_{\mathbf{k}})\varepsilon_{\mathbf{k}}/2}&\\
\times|n_{\mathbf{k}},\tilde{n}_{\mathbf{k}},\mathrm{in}\rangle \oplus|m_{\mathbf{k}},\tilde{m}_{\mathbf{k}},\mathrm{in}\rangle &,
\end{split}
\end{equation}
where $m_{\mathbf{k}}$ is the number of antiparticles corresponding to $b_{\mathbf{k},\mathrm{in}}$ and $\delta_{n_{\mathbf{k}},m_{\mathbf{k}}}$ presents charge conservation in every pure states. We denote particles with $\mathbf{a}$ and antiparticles with $\mathbf{b}$ where $\mathbf{\tilde{a}}$ and $\mathbf{\tilde{b}}$ are corresponding tilde particles. Because Eq.~\eqref{c+c} shows that a tilde particle is produced when a particle annihilates in thermal vacuum, the relation between tilde annihilation operator $\tilde{a}$ and $\tilde{b}$ has inverse results compared with Eq.~\eqref{ExternalBT}:
\begin{equation}\label{TildeExternalBT}
\left(
\begin{aligned}
& \tilde{b}_{\mathbf{\mathbf{k}},\mathrm{out}} \\
& \tilde{a}^{+}_{\mathbf{\mathbf{k}},\mathrm{out}}
\end{aligned}
\right) =
\left(
\begin{matrix}
\mu_{\mathbf{k}} & \nu^{*}_{\mathbf{k}} \\ \eta\nu_{\mathbf{k}} & \mu^{*}_{\mathbf{k}}
\end{matrix}
\right)
\left(
\begin{aligned}
& \tilde{b}_{\mathbf{\mathbf{k}},\mathrm{in}} \\
& \tilde{a}^{+}_{\mathbf{\mathbf{k}},\mathrm{in}}
\end{aligned}
\right).
\end{equation}
Then combining Eqs.~\eqref{ExternalBT}, \eqref{ThermalBT} and \eqref{TildeExternalBT}, the transformation of incoming particles under thermal rotation and external electromagnetic fields is
\begin{widetext}
\begin{equation}\label{TotalU}
\left(
\begin{aligned}
&a_{\mathbf{k},\mathrm{out}}(\theta) \\ &\tilde{a}^{+}_{\mathbf{k},\mathrm{out}}(\theta)
\\ &b^{+}_{\mathbf{k},\mathrm{out}}(\theta) \\ &\tilde{b}_{\mathbf{k},\mathrm{out}}(\theta)
\end{aligned}
\right)
=
\left(
\begin{matrix}
\mu_{\mathbf{k}}\mathrm{cos(h)}\theta_{\mathbf{k}} & -\mu_{\mathbf{k}}\mathrm{sin(h)}\theta_{\mathbf{k}}
&\nu^{*}_{\mathbf{k}}\mathrm{cos(h)}\theta_{\mathbf{k}} & -\nu^{*}_{\mathbf{k}}\mathrm{sin(h)}\theta_{\mathbf{k}} \\
\eta \mu^{*}_{\mathbf{k}}\mathrm{sin(h)}\theta_{\mathbf{k}} & \mu^{*}_{\mathbf{k}}\mathrm{cos(h)}\theta_{\mathbf{k}}
&\nu_{\mathbf{k}}\mathrm{sin(h)}\theta_{\mathbf{k}} & \eta \nu_{\mathbf{k}}\mathrm{cos(h)}\theta_{\mathbf{k}} \\
\eta \nu_{\mathbf{k}}\mathrm{cos(h)}\theta_{\mathbf{k}} & -\eta \nu_{\mathbf{k}}\mathrm{sin(h)}\theta_{\mathbf{k}}
&\mu^{*}_{\mathbf{k}}\mathrm{cos(h)}\theta_{\mathbf{k}} & -\mu^{*}_{\mathbf{k}}\mathrm{sin(h)}\theta_{\mathbf{k}} \\
\eta \nu^{*}_{\mathbf{k}}\mathrm{sin(h)}\theta_{\mathbf{k}} & \nu^{*}_{\mathbf{k}}\mathrm{cos(h)}\theta_{\mathbf{k}}
&\eta \mu_{\mathbf{k}}\mathrm{sin(h)}\theta_{\mathbf{k}} & \mu_{\mathbf{k}}\mathrm{cos(h)}\theta_{\mathbf{k}}
\end{matrix}
\right)
\left(
\begin{aligned}
&a_{\mathbf{k},\mathrm{in}} \\ &\tilde{a}^{+}_{\mathbf{k},\mathrm{in}}
\\ &b^{+}_{\mathbf{k},\mathrm{in}}\\&\tilde{b}_{\mathbf{k},\mathrm{in}}
\end{aligned}
\right).
\end{equation}
\end{widetext}
This transformation can be expressed as $T_{AB,\mathrm{out}}(\theta)=U(\theta)U(A^{ex})T_{AB,\mathrm{in}}U^{+}(A^{ex})U^{+}(\theta)$. Such that we can calculate pair production in thermal vacuum, when $t\rightarrow \infty$, the number is
\begin{equation}
\begin{split}
\mathcal{N}_{\mathbf{k},\mathrm{out}}(\beta_{q})
=&\langle0,\beta_{q},\mathrm{out}|a^{+}_{\mathbf{k},\mathrm{in}}a_{\mathbf{k},\mathrm{in}}|0,\beta_{q},\mathrm{out}\rangle \\
=&\langle0,\mathrm{in}|U^{+}(\theta)U^{+}(A^{ex})a^{+}_{\mathbf{k},\mathrm{in}}a_{\mathbf{k},\mathrm{in}}U(A^{ex})U(\theta)|0,\mathrm{in}\rangle \\
=&|\mu_{\mathbf{k}}|^{2}\mathrm{sin(h)}^{2}\theta_{\mathbf{k}}+|\nu_{\mathbf{k}}|^{2}\mathrm{cos(h)}^{2}\theta_{\mathbf{k}}.
\end{split}
\end{equation}
And there are particles condensing in initial thermal vacuum when $t\rightarrow -\infty$, where
\begin{equation}
\mathcal{N}_{\mathbf{k},\mathrm{in}}(\beta_{q})=\langle0,\beta_{q},\mathrm{in}|a^{+}_{\mathbf{k},\mathrm{in}}
a_{\mathbf{k},\mathrm{in}}|0,\beta_{q},\mathrm{in}\rangle=\mathrm{sin(h)}^{2}\theta_{\mathbf{k}}.
\end{equation}
So the number of created pairs is
\begin{equation}\label{NTq}
\Delta \mathcal{N}_{\mathbf{k}}(T_{q})=\mathcal{N}_{\mathbf{k},\mathrm{out}}(\beta_{q})-\mathcal{N}_{\mathbf{k},\mathrm{in}}(\beta_{q})=
\left\{
\begin{aligned}
&|\nu_{\mathbf{k}}|^{2}(1+2n^{B}_{\mathbf{k}}) \\
&|\nu_{\mathbf{k}}|^{2}(1-2n^{F}_{\mathbf{k}})
\end{aligned}
\right.
\end{equation}
Where $n^{B/F}_{\mathbf{k}}$ denotes either the Bose-Einstein distribution
$n^{B}_{\mathbf{k}}=1/(e^{\beta_{q} \varepsilon_{\mathbf{k}}}-1)$ for scalar QED or
$n^{F}_{\mathbf{k}}=1/(e^{\beta_{q} \varepsilon_{\mathbf{k}}}+1)$ for spinor QED.
This is consistent with \cite{Kim2009}. In Eq.~\eqref{NTq}, factor $|\nu_{\mathbf{k}}|^{2}$ response to the transformation under external fields
and $2|\nu_{\mathbf{k}}|^{2}\overline{n}^{B/F}_{\mathbf{k}}$ arises due to the present thermal charged particles. It conforms to the Bose-Einstein condensation for boson and the Pauli blocking for fermion.

When the density of particles and antiparticles is unequal or does not reach equilibrium density, the system can be in thermal but not in chemical equilibrium. We can redefine the distribution functions by a replacement $f^{\pm}=\lambda^{\pm}n^{\pm}$ as Ref.~\cite{Biro1993}, where superscript $\pm$ represent particles/antiparticles and $\lambda^{\pm}$ are fugacity factors for describing the deviation from chemical equilibrium, and add the chemical potential in density functions $n^{\pm}=1/(e^{\beta_{q}(\varepsilon\mp\mu)}+\eta)$. Taking these replacements into Eq.~\eqref{cossin} and Eq.~\eqref{TotalU}, we get
\begin{equation}
\Delta \mathcal{N}_{\mathbf{k}}(T_{q},\kappa)=|\nu_{\mathbf{k}}|^{2}(1-\eta(\lambda^{-}n^{-}_{\mathbf{k}}+\lambda^{+}n^{+}_{\mathbf{k}}))=
\left\{
\begin{aligned}
&|\nu_{\mathbf{k}}|^{2}(1+(\lambda^{-}n^{B,-}_{\mathbf{k}}+\lambda^{+}n^{B,+}_{\mathbf{k}})) \\
&|\nu_{\mathbf{k}}|^{2}(1-(\lambda^{-}n^{F,-}_{\mathbf{k}}+\lambda^{+}n^{F,+}_{\mathbf{k}}))
\end{aligned}
\right.
\end{equation}
When $\lambda^{+}=\lambda^{-}=1$, total charges are conservative, and the presence of the chemical potential enhances the thermal correction for $n^{-}_{\mathbf{k}}+n^{+}_{\mathbf{k}}>2n_{\mathbf{k}}$. The fugacity factors $\lambda^{\pm}$ depend on the ratio of the experimental to equilibrium density \cite{Thoma2009} and the non-ideal density $n^{\pm}$ also can be established as \cite{Gavrilov2008,Umezawa1995}.

Note that what we calculate is the number of incoming particles $a_{\mathrm{in}}$ which is not thermalized particles $a_{\mathrm{in}}(\theta)$.
Because the thermal transformation is equivalent to  Kubo-Martin-Schwinger condition, the thermal particles $a(\theta)$ satisfy corresponding condition, i.e. $\phi(t)=\pm\phi(t-i\beta)$. However, it is a basic condition in early imaginary-time effective lagrangian approaches at finite temperature, where the asymptotic particles are also assumed to meet this condition. So the number of thermalized particles is
\begin{equation}
\langle0,\beta_{q},\mathrm{out}|a^{+}_{\mathbf{k},\mathrm{in}}(\theta)
a_{\mathbf{k},\mathrm{in}}(\theta)|0,\beta_{q},\mathrm{out}\rangle=|\nu_{\mathbf{k}}|^{2}.
\end{equation}
This is just the result at zero temperature, so it may indirectly explain why some early discussions infer the absence of thermal corrections in pair production in one loop at finite temperature.

\subsection{\label{sec:r}Tree level correction of thermal photons}

Once charged pairs appear, thermal photons emerge as heat mediums in equilibrium QED systems. Inspired by paper \cite{Kohlfurst2014} which discusses mass shift when pairs produce in asymptotic periodic oscillating electromagnetic fields, we find these thermal Abelian gauge fields also change the asymptotic structure of charged particles, which cause inequivalent vacuum differed from original one. In this consideration, we assume that the total background fields consist of external fields and thermal fields, or more generally, any electromagnetic fields can be divided into:
\begin{gather}\label{FieldsEX+R}
A_{\mu}(x,T)=A_{\mu}^{ex}(x)+A^{r}_{\mu}(T_{r}),
\end{gather}
with an assumption
\begin{equation}\label{AAcommutation}
[A_{\mu}^{ex}(x),A^{r}_{\mu}(T)]=0.
\end{equation}
Which means that the two fields are both classical fields and only tree level Feynman diagrams are included. We assume that the external field is strong and thermal field is weak, so the two kinds of fields are absolutely separated and nonlinear terms are ignored.

Substituting Eq.~\eqref{FieldsEX+R} to Eq.~\eqref{KGDirac} and averaging it, all linear terms will be offset for thermal photons are isotropic.
Only $-q^{2}\langle A^{\gamma,\mu}A^{\gamma}_{\mu} \rangle$ is left and is equivalent to a mass shift in Eq.~\eqref{KGDirac}
\begin{equation}
m^{*2}=m^{2}-q^{2}\left< A^{\gamma,\mu}A^{\gamma}_{\mu} \right>.
\end{equation}
This $A^{r}(T)$ is assumed to be the sum of linear free photon fields, so the square term is proportional to invariant density $n_{\omega}/2\omega$.
After integrating, see Appendix A in detail, we obtain
\begin{equation}\label{massshift}
m^{*2}=m^{2}+\frac{q^{2}T^{2}}{12}.
\end{equation}
The square root of this shift part was also obtained in \cite{Loewe1992} and called effective mass of thermal photons \cite{Kibble1966}. Before the further discussions, we denote that $m^{*}_{+}=m^{*}$ and $m^{*2}_{-}=m^{2}-q^{2}T^{2}/12$. Next we will indicate that different asymptotic time conditions of thermal fields would lead to opposite thermal corrections.

Now we constrain a condition $A^{r}(t=\pm \infty)\neq0$ meaning thermal fields are always existing and can be treated as a static heat bath. Hence $|0,\mathrm{in};m^{*}_{+}\rangle$ and $|0,\mathrm{out};m^{*}_{+}\rangle$ are the dressed in-vacuum and out-vacuum. The $m^{*}_{+}$ in the kets denotes that naked mass $m$ of charged particles is directly substituted by dressed mass $m^{*}_{+}$. The corresponding number operator is $\mathcal{N}_{\mathbf{k}}(T_{r})=a^{+}_{\mathbf{k},\mathrm{in};m^{*}_{+}}a_{\mathbf{k},\mathrm{in};m^{*}_{+}}$, so what we calculate is the number of dressed particles
\begin{equation}
N_{\mathbf{k}}(T_{r})=\langle 0,\mathrm{out};m^{*}_{+}|\mathcal{N}_{\mathbf{k}}(T_{r})|0,\mathrm{out};m^{*}_{+}\rangle=|\nu_{\mathbf{k}}|^{2}_{m\rightarrow m^{*}_{+}}.
\end{equation}
It means the threshold has increased($E^{*}_{+}=2m^{*}_{+}c^{2}>2mc^{2}$) and more energy is demanded to counteract oscillating force turning into potential energy of dressed particles.

If we adopt a weaker condition $A^{r}(t=+\infty)=0$, the number operator should be a unaltered quantity which corresponds to final measurement. Similarly, the out-vacuum is supposed to in an equivalent measured Fock space \cite{Das1997}, or has the same mass term in Lagrangian compared with original one, and is denoted by $|0,\mathrm{out}\rangle$, corresponding the vacuum after evolution under $A^{ex}$ and $A^{r}$. In this scenario, as long as $A^{r}$ is turned off, the dressed particles roll into original particles and the vacuum restore into original vacuum, so the dressing thermal photons turn into energy of produced particles that enhance pair production. Next we verify it in an evolutive approach instead of standard background method. The out-vacuum is
\begin{equation}\label{U+Uin}
\begin{split}
|0,\mathrm{out}\rangle
&\approx e^{i\int dt\int d^{3}x \overline{\mathcal{L}}(A^{ex},A^{r})-\mathcal{L}(A^{ex})}e^{i\int d^{4}x \mathcal{L}(A^{ex})}|0,\mathrm{in};m^{*}_{-}\rangle \\
&=U(A^{r}|A^{ex})U(A^{ex})|0,\mathrm{in};m^{*}_{-}\rangle.
\end{split}
\end{equation}
In the first line, we use Eq.~\eqref{massshift} and an assumption that $A^{r}(T_{r})$ is small. The $\overline{\mathcal{L}}$ is Lagrangian density under external
fields $A^{ex}$ and thermal fields $A^{r}$, where the overline means that the $A^{r}$ has been averaged and will vanish when $t\rightarrow \infty$.
The $\mathcal{L}(A^{ex})$ is just a normal Lagrangian density. Their detailed expressions depend on what vacuum they act on, and we do not need to
give a concrete expression here. But we divide the evolution operator into two parts: the $U(A^{ex})$ which undertakes main evolution of vacuum and $U(A^{r}|A^{ex})$
which changes the asymptotic structure of vacuum. It is obvious that
\begin{equation}
\begin{split}
&U(A^{ex})|0,\mathrm{in};m^{*}_{-}\rangle =|0,\mathrm{out};m^{*}_{-}\rangle, \\
&U(A^{r}|A^{ex})|0,\mathrm{out};m^{*}_{-}\rangle =|0,\mathrm{out}\rangle.
\end{split}
\end{equation}
And the commutative relation $[U(A^{r}|A^{ex}),U(A^{ex})]=0$ holds due to Eq.~\eqref{AAcommutation}, therefore the evolutionary order does not affect the final result. Furthermore, because the result of $\overline{\mathcal{L}}(A^{ex},A^{r})-\mathcal{L}(A^{ex})$ is a mass term, $U(A^{r}|A^{ex})$ is an unitary operator. Applying above analysis, we get that
\begin{equation}\label{UAU}
U^{+}(A^{r}|A^{ex}) a_{\mathbf{k},\mathrm{in}} U(A^{r}|A^{ex})~U^{+}(A^{r}|A^{ex}) |0,\mathrm{in}\rangle
=a_{\mathbf{k},\mathrm{in};m^{*}_{-}} |0,\mathrm{in};m^{*}_{-}\rangle = 0 .
\end{equation}
Now with above relations, we can obtain the number of produced pairs in this scenario as
\begin{equation}\label{VNR}
\begin{split}
N_{\mathbf{k}}(T_{r})
=&\langle0,\mathrm{out}|a^{+}_{\mathbf{k},\mathrm{in}}a_{\mathbf{k},\mathrm{in}}|0,\mathrm{out}\rangle \\
=&\langle0,\mathrm{in};m^{*}_{-}|U^{+}(A^{ex})U^{+}(A^{r}|A^{ex})a^{+}_{\mathbf{k},\mathrm{in}}a_{\mathbf{k},\mathrm{in}}
 U(A^{r}|A^{ex})U(A^{ex})|0,\mathrm{in};m^{*}_{-}\rangle \\
=&\langle0,\mathrm{out};m^{*}_{-}|a^{+}_{\mathbf{k},\mathrm{in};m^{*}_{-}}a_{\mathbf{k},\mathrm{in};m^{*}_{-}}|0,\mathrm{out};m^{*}_{-}\rangle \\
=&|\nu_{\mathbf{k}}|^{2}_{m\rightarrow m^{*}_{-}} .
\end{split}
\end{equation}
In this result, Eq.~\eqref{U+Uin} and Eq.~\eqref{UAU} are used. Note that when $U(A^{ex})$ acts on vacuum, the result depends on the asymptotic structure of the vacuum, so does $U(\theta)$. This result means in order to take tree level correction of thermal photons into pair production, we only need to do a mass shift in Bogoliubov coefficient, so the threshold has gone down($E^{*}_{-}=2m^{*}_{-}c^{2}<2mc^{2}$), which verifies our guess that dressing thermal photons enhance pair production. In subsequent discussions, we will adopt this approach.

More generally, stable equilibrium QED systems should contain both thermal charged particles and thermal photons. Following above approaches, the thermal
out-vacuum also is an equivalent measured thermal vacuum
\begin{equation}
|0,0,\beta_{q},\mathrm{out}\rangle=U(A^{r}|A^{ex})U(A^{ex})U(\theta)|0,0,\mathrm{in};m^{*}_{-}\rangle.
\end{equation}
Then we do similar calculations as Eq.~\eqref{VNR}, and pair production in this situation is obtained as
\begin{equation}\label{NTqr}
N_{\mathbf{k}}(T_{q},T_{r})=[|\nu_{\mathbf{k}}|^{2}(1\pm n^{B/F}_{\mathbf{k}})]_{m\rightarrow m^{*}_{-}}.
\end{equation}
This result is just Eq.~\eqref{NTq} with a negative mass shift, which means that the tunnelling particles and thermal particles all have absorbed thermal photons. A remark is that the $\overline{n}^{F/B}_{\mathbf{k}}$ in Eq.~\eqref{NTqr} depends on ont only charged particles temperature $T_{q}$ in Boltzmann factor
$e^{\varepsilon_{\mathbf{k}}/T_{q}}$, but also thermal photons temperature $T_{r}$ in $\varepsilon_{\mathbf{k}}=\sqrt{m^{2}-q^{2}T^{2}_{r}/12+\mathbf{k}^{2}}.$
Here we ignored any nonlinear and high terms due to the assumation that $T_{r}$ is small $q^{2}T^{2}/12 \ll m^{2}$ that $m^{*2}_{-}$ is always positive.

Up to now, we have not used any particular external fields, and it is for the first time to obtain formal results of Schwinger pair production at finite temperature with
counting tree level correction of real thermal photons, compared with previous  results\cite{Gould2017,Monin2010,Brown2018,Draper2018,Medina2017,Gould2019,Dittrich1979,Gies1999oneloop,Elmfors1995}.

\section{\label{sec:II}Constant electric field}
In this section, we apply the formal results above to an external constant electric field. Before considering finite temperature,
we review the solutions of Eq.~\eqref{KGDirac} in a constant electric field \cite{Kim2008}. In the time-dependent gauge $A_{z}=-Et$,
Eq.~\eqref{KGDirac} takes the form
\begin{equation}\label{KGDiracE}
\left[\partial^{2}_{t}+m^{2}+\mathbf{k}^{2}_{\bot}+(k_{z}-qEt)^{2}+2isqE\right]\phi_{\omega,\mathbf{k}}(t)=0.
\end{equation}
When $t=-\infty$, the solution of the parabolic cylinder function is expressed as
\begin{equation}
\phi_{\omega,\mathbf{k}}(t)=D_{p}(z).
\end{equation}
And when $t=\infty$, it turns to
\begin{equation}
D_{p}(z)=e^{-ip\pi}D_{p}(-z)+\frac{\sqrt{2\pi}}{\Gamma(-P)}e^{-i(p+1)\pi/2}D_{-p-1}(iz),
\end{equation}
where
\begin{equation}
\begin{split}
z&=\sqrt{\frac{2}{qE}}e^{i\pi/4}(k_{z}-qEt),\\
p&=-\frac{1}{2}-i\frac{m^{2}+\mathbf{k}^{2}_{\bot}+2isqE}{2qE}.
\end{split}
\end{equation}
The Bogoliubov coefficients are
\begin{equation}
\mu_{\mathbf{k}}=\frac{\sqrt{2\pi}}{\Gamma(-p)}e^{-i(p+1)\pi/2},\quad \nu_{\mathbf{k}}=e^{ip\pi}.
\end{equation}
So the total number of produced pair is
\begin{equation}\label{NumberFinalE}
\begin{split}
N(E,T_{q}=T_{r}=0)&= \sum_{\mathbf{k}} |\nu_{\mathbf{k}}|^{2}\\
                  &= \frac{qE}{4\pi(1-|s|)} \int \frac{d^{2}k_{\perp}}{(2\pi)^{2}} e^{-\frac{\pi(m^{2}+k_{\perp}^{2})}{qE}}\\
                  &= \frac{q^{2}E^{2}}{16(1-|s|)\pi^{3}} e^{-\frac{\pi m^{2}}{qE}},
\end{split}
\end{equation}
where, $\frac{qE}{4\pi(1-|s|))}$ is the number of states along the $z$-direction and Eq.~\eqref{NumberFinalE} is consistent with Schwinger's result \cite{Schwinger1951}
when $|s|=1/2$ for spinor.

In the following, thermal gauge fields are assumed to be turned off when $t\rightarrow\infty$ as long as they are considered. Meanwhile, we assume that all systems are in equilibrium initially although an external constant field is not good to get asymptotically free states. The time scale of pair production is also assumed to be much shorter than the time scale of thermal dissipation of produced pairs, so the back-reactions are neglected, where low temperature and low density limit also been assumed to accord with it. In a word, systems are thermalized initially but process of producing pairs are dynamic.

\subsection{\label{sec:Er}$T_{q}=0, T_{r}\neq0$}
Now, we consider a pure thermal photon system with a constant electric field, where the low temperature limit $q^{2}T^{2}_{r}/12\ll m^{2}$ is reasonable since no real particles exist initially due to low quantum fluctuation. But the tunneling particles are also dressed by thermal photons. In this way, the number of produced pairs is obtained by substituting $m^{2}$ in Eq.~\eqref{NumberFinalE} with $m^{2}-q^{2}T^{2}_{r}/12$ according to Eq.~\eqref{VNR}
\begin{equation}
N(E,T_{r})=\frac{(qE)^{2}}{16(1-|s|)\pi^{3}} e^{\frac{-\pi m^{2}}{qE}}e^{\frac{\pi qT^{2}_{r}}{12E}}.
\end{equation}
At small $T_{r}$, this result has a series expanded form
\begin{equation}
\begin{split}
N(E,T_{r})=&N(E,T_{q}=T_{r}=0) \\
&\times \left(1+\frac{\pi qT^{2}_{r}}{12E}+\frac{1}{2}(\frac{\pi q}{12E})^{2}T^{4}_{r}+\cdots \right).
\end{split}
\end{equation}

The exponent $T^{2}_{r}/E$ means that the more stronger electric field is, the less effect fixed thermal photons cause.
In Ref.~\cite{Gies2000twoloop}, thermal photons are considered by two-loop calculation and lowest order thermal correction term contains $T^{4}$,
but we get $T^{2}$. This is because that thermal fields are treated as external current in our discussion, while those are internal in \cite{Gies2000twoloop} where more double vertexes in diagram expansion exist.

\subsection{\label{sec:Eq}$T_{q}\neq0, T_{r}=0$}

In this physical situation, the temperature $T_{q}$ of thermal charged particles is not supposed to be high enough to produce real photons. Then, with a constant electric field, the result is obtained directly by combining Eq.~\eqref{NTq} and Eq.~\eqref{NumberFinalE}, that is
\begin{equation}\label{NumberFinalQE}
\begin{split}
 N(E,T_{q})  =& \frac{qE}{4\pi(1-|s|)} \int \frac{d^{2}k_{\perp}}{(2\pi)^{2}} e^{-\frac{\pi(m^{2}+k_{\perp}^{2})}{qE}}  \\
                               & \times \left( 1 - \frac{2\eta}{e^{ \sqrt{m^{2}+k_{\perp}^{2}}/T_{q}} +\eta} \right).
\end{split}
\end{equation}
Here, the longitudinal momentum of initial particles is assumed to be zero, where an extra force is constructed to
neutralize the electric field force. It returns to \eqref{NumberFinalE} when $T_{q}\rightarrow0$, but tends to either infinity for bosons
or zero for fermions when $T_{q}\rightarrow\infty$. When return to $T_{q}<<m$, the number of produced pairs \eqref{NumberFinalQE} has an approximate form
\begin{equation}
\begin{split}
N(E,T_{q})=&\frac{(qE)^{2}}{16(1-|s|)\pi^{3}}e^{-\pi m^{2}/(qE)}\\
                             &\times \left( 1-\eta\frac{qE}{4\pi^{2}}e^{-\pi m^{2}/(qE)- m/T_{q}} \right).
\end{split}
\end{equation}

\subsection{\label{sec:Eqr}$T_{q}=T_{r}\neq0$}

Finally, we assume that the initial systems are in absolute equilibrium where thermal photons and charged particles have same temperature $T_{q}=T_{r}=T$ with an external constant electric field: the system is actually an ideal particle-antiparticle-photon gas.
Combining Eq.~\eqref{NumberFinalQE} and Eq.~\eqref{NTqr}, and the number of produced pairs is obtained as
\begin{equation}
\begin{split}
N(E,T_{q}=T_{r}=T)=& \frac{qE}{4\pi(1-|s|)} \int \frac{d^{2}k_{\perp}}{(2\pi)^{2}}
                                 e^{-\frac{\pi(m^{2}+k_{\perp}^{2})}{qE}+\frac{\pi qT^{2}}{12E}}\\
                               & \times \left(1 - \frac{2\eta}{e^{\beta \sqrt{m^{2}-q^{2}T^{2}/12+k_{\perp}^{2}}} +\eta}\right),
\end{split}
\end{equation}
which has an approximate form at small $T$
\begin{equation}
\begin{split}
N(E,T_{q}=T_{r}=T\ll m)=&\frac{qE}{4(1-|s|)\pi}e^{\frac{-\pi m^{2}}{qE}}e^{\frac{\pi qT^{2}}{12E}} \\
&\times \left( 1-\eta \frac{qE}{4\pi^{2}}e^{-\sqrt{m^{2}/T^{2}-q^{2}/12}} \right).
\end{split}
\end{equation}

So far, we have not use any limit on electric field $E$, but low temperature limit for $T_{r}$ is necessary for our linear assumption in Sec.~II.~C. In fact, we expect the external field is much stronger than temperature so that the thermal dissipation and back-reactions can be neglected.

\section{Discussion}\label{discussion}

In this paper, to calculate Schwinger pair production at finite temperature, we generalize the evolution operator method of TFD to generic thermal QED systems that contain not only thermal charged particles but also photons. Then we get thermal transformation relation of outgoing operators and incoming operators under external fields. What we have done on the basis of Eq.~\eqref{KGDirac} is similar to Sauter's calculation on Dirac equation \cite{Sauter1931}, except for the part where we add thermal distribution. Then we use an effective substitute $m^{*2}_{+}=m^{2}+q^{2}T^{2}/12$ to represent the tree level correction of thermal photons, but that has an inverse substitution when the turning off of thermal fields is considered. Lastly, applying thermal average approach $\langle \mathrm{out- vacuum}|a^{+}_{\mathrm{in}}a_{\mathrm{in}}|\mathrm{out-vacuum}\rangle$, we get QED pair production in thermal systems. In our results, $|\nu_{\mathbf{k}}|^{2}$ corresponds to decaying part of incoming states under external field, and density $n^{B/F}_{\mathbf{k}}$ arises for initial thermal charged particles and $m^{*2}_{+}$ represents effective mass of both thermal particles and tunneling particles dressed by thermal photons. In an external constant electric field, the precise integral results and the approximate polynomial results are obtained and they both recover Schwinger's result at $T=0$.

Back to our results, we haven't used any particular fields but only the Bogoliubov coefficient $\nu_{\mathbf{k}}$ in our formal results,
so it can be applied to arbitrary fields as long as we obtain the coefficient. There are no poles or infinity to be renormalized, since the number of produced pairs is an observable quantity. If charged particles or photons are not in ideal equilibrium initially, it can be completed by using a corresponding distribution function to replace ideal one as we have down in Sec.~II.~B. It also can be generalized to thermal Bethe-Heitler process $(\gamma Z\rightarrow e^{+}e^{-}Z)$ and thermal trident mechanism $(e^{-}Z\rightarrow e^{-}e^{+}e^{-}Z)$ where thermal massive charged particle $Z$ exists \cite{Maltsev2019}, and classical Coulomb field can be taken into background field as equivalent approach,
but the regions need to be distinguished for the Debye shielding. So it may be helpful in relativistic plasma and heavy ion collision.

On the experimental side, it is proposed to observe thermal Schwinger pair production by a thermal bath of photons in a constant electric field \cite{Rose2013},
and our results in a constant $E$ supplement neglected correction. What more is that our formal results can be applied to general laser fields by
combining numerical simulation \cite{Krekora2005}. In addition, momentum spectrum distribution of production particles is given in our results,
and spatial spectral distribution also can be obtained by Fourier transform. These results are available by comparisons with that in possible future experiments.

Again, our results mainly work in low temperature limit $T_{r}\ll m$, where back-reactions and dissipations have not be considered. With temperature increasing, full equilibrium description is more appropriate, more detailed summary and review one can refer to \cite{Gould2019}. By the way to generalize temperature to arbitrary scale, the effective Lagrangian approaches are more proper, which is beyond the present study and is still an open topic problem for future research.
\begin{appendix}

\section{effective mass of thermal photons}\label{APPENDIX}

In Sec.~II.~B, we illustrate the use of mass shift with thermal photons, that is
\begin{equation}\label{Massshift}
m^{*2}=m^{2}+\frac{q^{2}T^{2}}{12},
\end{equation}
where $q^{2}T^{2}/12,$ is square of real thermal photon effective mass, which is different from the effective photon mass, $m_{eff}=qT/3$ which is calculated in one-loop polarization \cite{Weldon1982}. Now, we derive it with semiclassical approaches, begin with
\begin{equation}\label{MassshiftA}
m^{*2}=m^{2}-q^{2} \langle A^{r,\mu}A^{r}_{\mu} \rangle,
\end{equation}
where $A_{r}$ is the linear sum of all thermal photons and $\langle \rangle$ indicates averaging over time and space. We assume
that all thermal photons are free and satisfy $\langle A^{f,\mu}A^{f}_{\mu}\rangle=-1/2\omega$, and integrate this relation by
\begin{equation}\label{A2}
\langle A^{r,\mu}A^{r}_{\mu} \rangle=-\int^{\infty}_{0}d\omega\frac{n(\omega)}{2\omega}.
\end{equation}
Where $n(\omega)/2w$ is the invariant density and $n(\omega)$ equal to
\begin{equation}\label{A3}
n(\omega)=\frac{g(\omega)}{e^{\beta \omega}-1}=\frac{\omega^{2}}{\pi^{2}(e^{\beta \omega}-1)},
\end{equation}
where $g(\omega)$ is the energy density and it's value is inserted directly here. Combining Eqs.~\eqref{MassshiftA}, \eqref{A2} and \eqref{A3},
we obtain Eq.~\eqref{Massshift}.

\end{appendix}

\begin{acknowledgments}
We thank Dr. Z. L. Li for helpful discussions and M. Ababekri for his reading of the paper and modification to language. This work was supported by the National Natural Science Foundation of China (NSFC) under Grants No. 11875007 and No. 11935008.
\end{acknowledgments}


\begin{thebibliography}{99}\suppressfloats
\bibitem{Schwinger1951}
J. S. Schwinger, Phys. Rev. 82, 664 (1951).

\bibitem{Gavrilov2019}
S. P. Gavrilov, D. M. Gitman and A. A. Shishmarev, Phys. Rev. D. 99, 116014 (2019).

\bibitem{Dunne2014}
G. Dunne, Eur. Phys. J. Special Topics 223, 1055 (2014).

\bibitem{Piazza2012}
A. DiPiazza, C. Muller, K. Z. Hatsagortsyan and C. H. Keitel, Rev. Mod. Phys. 84. 1177 (2012).

\bibitem{Gould2017}
O. Gould and A. Rajantie, Phys. Rev. D 96, 076002 (2017).

\bibitem{Monin2010}
A. Monin and M. B. Voloshin, Phys. Rev. D 81, 025001 (2010).

\bibitem{Brown2018}
A. R. Brown, Phys. Rev. D 98, 036008 (2018).

\bibitem{Draper2018}
P. Draper, Phys. Rev. D 98, 125014 (2018).

\bibitem{Medina2017}
L. Medina and M. C. Ogilvie, Phys. Rev. D 95, 056006 (2017).

\bibitem{Korwar2018}
M. Korwar and A. M. Thalapillil, Phys. Rev. D 98, 076016 (2018).

\bibitem{Torgrimsson2019}
G. Torgrimsson, Phys. Rev. D 99, 096007 (2019).

\bibitem{Gould2019}
O. Gould, S. Mangles, A. Rajantie, S. Rose and C. Xie, Phys. Rev. A 99, 052120 (2019).

\bibitem{Gelis2016}
F. Gelis and N. Tanji, Prog. Part. Nucl. Phys. 87, 1 (2016).

\bibitem{Dittrich1979}
W. Dittrich, Phys. Rev. D 19, 2385 (1979).

\bibitem{Gies1999oneloop}
H. Gies, Phys. Rev. D 60, 105002 (1999).

\bibitem{Elmfors1995}
P. Elmfors and B. S. Skagerstam, Phys. Lett. B 348, 141 (1995).

\bibitem{Gies2000twoloop}
H. Gies, Phys. Rev. D 61, 085021 (2000).

\bibitem{Thoma2009}
M. H. Thoma, Eur. Phys. J. D 55, 271 (2009).

\bibitem{Kim2008}
S. P. Kim and H. K. Lee, Phys. Rev. D 76, 125002 (2007); Phys. Rev. D 78, 105013 (2008).

\bibitem{Gavrilov2008}
S.P. Gavrilov and D.M. Gitman, Phys. Rev. D 78, 045017 (2008).

\bibitem{Gavrilov2008matrix}
S.P. Gavrilov, D.M. Gitman, and J.L. Tomazelli, Nucl. Phys. B 795, 645 (2008).

\bibitem{Krekora2006}
P. Krekora, Q. Su and R. Grobe, Phys. Rev. A 73, 022114 (2006).

\bibitem{Kim2009}
S. P. Kim, H. K. Lee and Y. Yoon, Phys. Rev. D 79, 045024 (2009).

\bibitem{Umezawa1995}
H. Umezawa, Advanced field theory: Micro, macro, and thermal physics (AIP, 1995).

\bibitem{K2010}
S. P. Kim, H. K. Lee and Y. Yoon, Phys. Rev. D 82, 025016 (2010).

\bibitem{Rose2013}
S. J. Rose, High Energy Density Phys. 9, 480 (2013).

\bibitem{Gies2017}
H. Gies and F. Karbstein, JHEP 03, 108 (2017).

\bibitem{Braun1999}
J. W. Braun, Q. Su and R. Grobe, Phys. Rev. A 59, 604 (1999).

\bibitem{Biro1993}
T. S. Biro, E. van Doorn, B. Muller, M. H. Thoma and X. N.
Wang, Phys. Rev. C 48, 1275 (1993).

\bibitem{Blinne2019}
A. Blinne, H. Gies, F. Karbstein, C. Kohlfurst and M. Zepf, Phys. Rev. D 99, 016006 (2019).

\bibitem{Affleck1981}
I. Affleck, Phys. Rev. Lett. 46, 388 (1981).

\bibitem{Sauter1931}
F. Sauter, Z. Phys. 69, 742 (1931).

\bibitem{Ojima1981}
I. Ojima, Ann, Phys. 137, 1 (1981).

\bibitem{Krekora2005}
P. Krekora, K. Cooley, Q. Su and R. Grobe, Phys. Rev. Lett. 95, 070403 (2005).

\bibitem{Loewe1992}
M. Loewe and J. C. Rojas, Phys. Rev. D 46, 2689 (1992).

\bibitem{Kibble1966}
T. W. B. Kibble, Phys. Rev. 150, 1060 (1966).

\bibitem{Kohlfurst2014}
C. Kohlfurst, H. Gies and R. Alkofer, Phys. Rev. Lett. 112, 050402 (2014).

\bibitem{Maltsev2019}
I. A. Maltsev, V. M. Shabaev, R. V. Popov, Y. S. Kozhedub, G. Plunien, X. Ma, T. Stohlker and D. A. Tumakov, Phys. Rev. Lett. 123, 113401 (2019).

\bibitem{Das1997}
A. Das, Finite Temperature Field Theory (World Scientific, Singapore, 1997).

\bibitem{Weldon1982}
H. A. Weldon, Phys. Rev. D 26, 1394 (1982).
\end{thebibliography}
\end{document}